\documentclass[12pt]{article}

\usepackage{graphics, color}
\usepackage{graphicx}
\usepackage{amssymb}
\usepackage{amsmath}
\usepackage{slashed}
\usepackage{cite}
\pdfoutput=1

\def\gsim{\, \rlap{$>$}{\lower 1.1ex\hbox{$\sim$}}\,}
\def\lsim{\, \rlap{$<$}{\lower 1.1ex\hbox{$\sim$}}\,}

\def\F{{{\cal F}}}

\def\Z{{{\mathbb Z}}}

\def\Re{{{\frak{Re}}}}

\def\CE{{\cal E}}
\def\CL{{\mathcal{L}}}
\def\CO{{\cal O}}

\def\tr{{\rm tr}}

 \newcommand{\be}{\begin{equation}}
\newcommand{\ee}{\end{equation}}
 \newcommand{\bal}{\begin{align}}
 \newcommand{\eal}{\end{align}}
\newcommand{\ben}{\begin{equation*}}
\newcommand{\een}{\end{equation*}}
\newcommand{\bea}{\begin{eqnarray}}
\newcommand{\eea}{\end{eqnarray}}
\newcommand{\bean}{\begin{eqnarray*}}
\newcommand{\eean}{\end{eqnarray*}}
\newcommand{\bes}{\begin{subequations}}
\newcommand{\ees}{\end{subequations}}

\def\p{\partial}
\def\d{\partial}

\def\eps{{\epsilon}}

\def\CE{{{\cal E}}}
\def\RR{{\mathbb R}}
\def\cpn{\mathbb{CP}^N}

\def\half{\frac{1}{2}}

\def\tF{{\tilde F}}
\def\te{{\tilde e}}
\def\tA{{\tilde A}}
\def\tR{{\tilde R}}
\def\tG{{\tilde G}}
\def\tE{{\tilde E}}
\def\cf{{\it cf.}}

\usepackage{epsfig}

\begin{document}


\begin{titlepage}

\begin{flushright}
BRX-TH-644
\end{flushright}

\bigskip
\bigskip\bigskip\bigskip
\centerline{\Large $\theta$-angle monodromy in two dimensions}
\bigskip\bigskip\bigskip

\centerline{{\bf Albion Lawrence}}
\medskip
\centerline{\em Martin Fisher School of Physics, Brandeis University}
\centerline{\em MS 057, 415 South Street, Waltham, MA 02454}
\bigskip
\bigskip\bigskip


\begin{abstract}

\noindent "$\theta$-angle monodromy" occurs when a theory possesses a landscape of metastable vacua which reshuffle as one shifts a periodic coupling $\theta$ by a single period.  "Axion monodromy" models arise when this parameter is promoted to a dynamical pseudoscalar field.  This paper studies the phenomenon in two-dimensional gauge theories which possess a $U(1)$ factor at low energies: the massive Schwinger and gauged massive Thirring models, the $U(N)$ 't Hooft model, and the ${\mathbb CP}^N$ model. In all of these models, the energy dependence of a given metastable false vacuum deviates significantly from quadratic dependence on $\theta$ just as the branch becomes completely unstable (distinct from some four-dimensional axion monodromy models). In the Schwinger, Thirring, and 't Hooft models, the meson masses decrease as a function of $\theta$.  In the $U(N)$ models, the landscape is enriched by sectors with nonabelian $\theta$ terms.  In the ${\mathbb CP}^N$ model, we compute the effective action and the size of the mass gap  is computed along a metastable branch.   

\end{abstract}
\end{titlepage}
\baselineskip = 17pt
\setcounter{footnote}{0}

\section{Introduction}

In four-dimensional quantum field theories, the potential energy for a periodic scalar $\phi$ such as an axion is often taken to be a bounded periodic function, {\it e.g.} $V(\phi) = \Lambda^4 \cos (\phi/f)$.  Such potentials can be generated by instanton effects; the periodicity $\phi \to \phi + 2\pi f$ protects the theory from perturbative corrections of the form $\phi^n$.

\begin{figure}[ht!]
\begin{center}
\includegraphics[scale=.85]{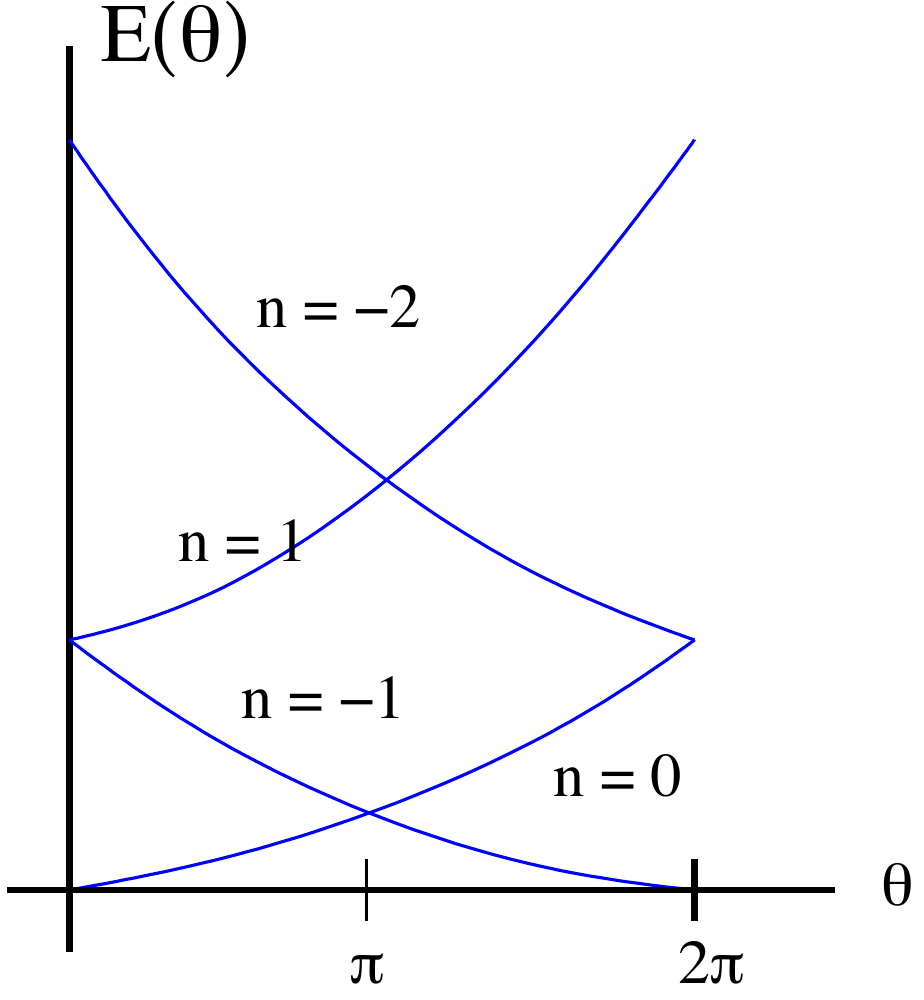}\end{center}
\vspace{-.5cm}
\caption{\label{monofig}The potential energy for a periodic scalar $\phi = f \theta$.  For $\theta$ fixed, the lowest energy branch corresponds to the ground state, and the higher-energy branches are metastable.  The theory has a first-order quantum phase transition at $\theta = 2\pi (n + \half)$.}
\end{figure}

This is not the only option for a periodic scalar.  The theory may be invariant under shifts $\phi \to\phi + 2\pi f$, but the energy spectrum can shift, so that the potential energy curves appear as in Figure \ref{monofig}.  When the spatial volume is infinite, there is a first-order quantum phase transition at the point $\phi = \pi f$ where the levels cross.\footnote{At finite volume, in interacting theories, we expect the level crossings to split and the energy spectrum to break up into bands, as discussed in \cite{Kaloper:2011jz}.}    This phenomenon is known to occur in large-N QCD \cite{Witten:1979vv,Witten:1980sp}.  We will dub such a phenomenon "axion monodromy" (after \cite{McAllister:2008hb}), or "theta angle monodromy" in the case that $\phi/f$ couples as a theta term and we freeze its dynamics.

\begin{figure}[ht!]
\begin{center}
\includegraphics[scale=.5]{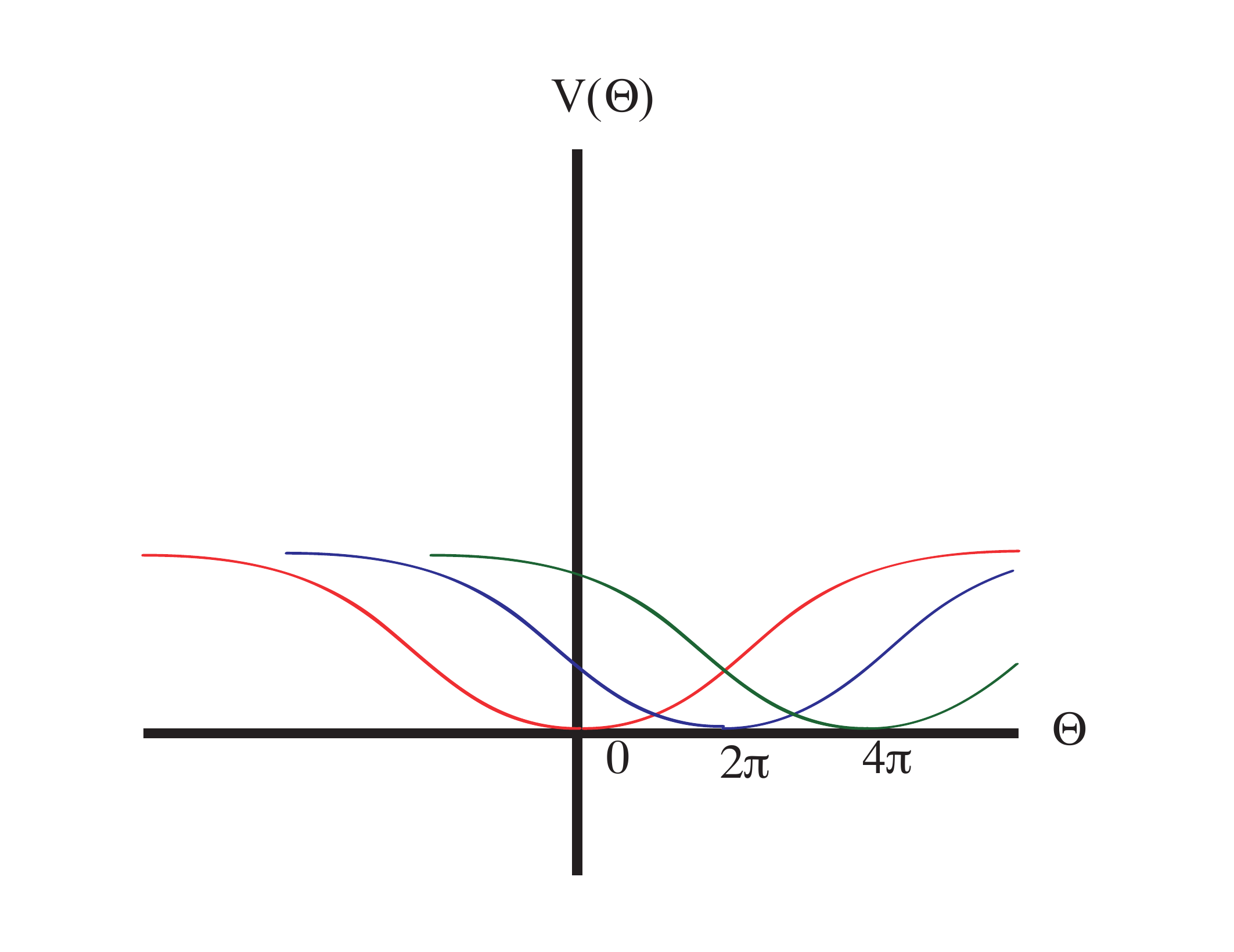}\end{center}
\vspace{-.5cm}
\caption{\label{Branches}The energy as a function of the $\theta$ term for the large-N theory studied in \cite{Dubovsky:2011tu}, for three branches of the theory. The lines with larger energy correspond to metastable vacua; far out along a given branch, the metastable vacua flatten out and become increasingly unstable to decay to the lower branches.}
\end{figure}

Monodromy in field space leads to an interesting class of models of inflation in string and field theory \cite{Silverstein:2008sg,McAllister:2008hb,Kaloper:2008fb,Berg:2009tg,Kaloper:2011jz, Dubovsky:2011tu}.  Most of these models have potentials which are quadratic in the axion close to the minimum of the branch, and flatten out far along the branch \cite{Silverstein:2008sg,McAllister:2008hb,Dong:2010in,Dubovsky:2011tu}, as shown in Fig. \ref{Branches}. Inflation takes place in this flattened regime.  In the theory studied in \cite{Dubovsky:2011tu}, the flattening appears related to a lowering of the mass gap of the confining gauge theory as a function of $\phi$, and is generated by the same dynamics that generates the monodromy. Refs. \cite{Kaloper:2008fb,Kaloper:2011jz,Dubovsky:2011tu}\ also studied the nonperturbative instability of the higher energy branches.  In the strongly-coupled large N theory described in \cite{Dubovsky:2011tu}\ the branches become completely unstable deep in the "flat" regime to decaying to a lower branch; again, this arises from nonperturbative gauge dynamics. In the axion-four form theory described in \cite{Kaloper:2008fb,Kaloper:2011jz}, this separation requires that the axionic domain wall tension be larger than the UV scale governing irrelevant operators of the theory.

This paper arose from an attempt to better understand the theories studied in \cite{Kaloper:2011jz,Dubovsky:2011tu}\ by studying two-dimensional models with a theta term and theta angle monodromy. We will investigate the massive Schwinger model, the gauged massive Thirring model, $U(N)$ gauge theories coupled to fundamental matter, and the large-N ${\mathbb CP}^N$ model.  In the first three models we will set the gauge coupling to be smaller than the fermion mass, so that there is a tower of metastable states as shown in Figure 1.  In the ${\mathbb CP}^N$ model, the low-energy theory is an abelian vector field coupled to charged matter, with the dynamically generated gauge coupling $\CO(1/N)$ times the dynamically generated mass of the charged particles.  In all of these cases, we will find that, unlike the four-dimensional model studied in \cite{Dubovsky:2011tu}, the onset of $\CO(\theta^4)$ corrections to the quadratic behavior of the energy $\CE(\theta) \sim \theta^2$ occurs precisely when the branch becomes unstable.  At present I do not have a really satisfying explanation for this; it is possible that it is related to the fact that in two dimensions,  the theta term couples to an abelian factor of the gauge group. I should also note corrections to $\CE(\theta)\sim\theta^2$ are possible if additional neutral degrees of freedom couple to the gauge sector, for example via a field-dependent gauge coupling; such couplings were shown to lead to flattening in \cite{Dong:2010in}.

The basic story for the three models of charged fermions is that the $\theta$-dependent dynamics of all of these theories at sufficiently low energies is well-described by the Sine-Gordon model:
\be\label{eq:generalform}
	L = K \left[ \half (\p\phi)^2 + \mu^2 \cos\phi - \frac{(\phi + \theta)}{2\pi} \tF_{01} -\frac{1}{2\te^2} \tF_{01}^2\right]
\ee
where $\te,\tF_{01}$ are suitably rescaled $U(1)$ gauge coupling and electric field.  The resulting potential energy is shown in Figure 3.  $K$ is $\sim \CO(1)$ for the Schwinger model, is proportional to the four-fermion coupling for the Thirring model, and is proportional to the rank $N$ of the $U(N)$ gauge group for the 't Hooft model.  Thus these latter examples have a semiclassical limit $K\to\infty$.  Let us consider adiabatically increasing $\theta$ over many periods.  In this case, we will describe $\theta$ as living on $\RR$ (the covering space of $S^1$), and the highly metastable states as lying at "large $\theta$".  If one begins in the true ground state and adiabatically increases $\theta$, a given vacuum becomes a metastable false vacuum.  The mass of scalar fluctuations about the minimum of the false vacuum (this corresponds to a meson mass) decreases with $\theta$.  When $\te^2 \theta \sim \tF_{01} > mu^2$, the false vacuum becomes unstable.  When $K \gg 1$, and $\te^2 \theta < \mu^2$, corrections to the quadratic $\theta$-dependence of the energy of the false vacua arise from integrating out $\phi$ classically, which leads to corrections of the form $(\te^2\theta/\mu^2)^k$.  These become important just as the false vacuum becomes unstable. The same phenomenon occurs in the Schwinger model, as can be seen by integrating out the fermions directly. The ${\mathbb CP}^N$ model at low energies is essentially a multiflavor bosonic version of the Schwinger model -- as we will see, the number of flavors {\it reduces}\ by $\CO(1/\ln N)$ the value of $\theta$ at which a metastable branch becomes unstable.

\begin{figure}[ht!]
\begin{center}
\includegraphics[scale=.6]{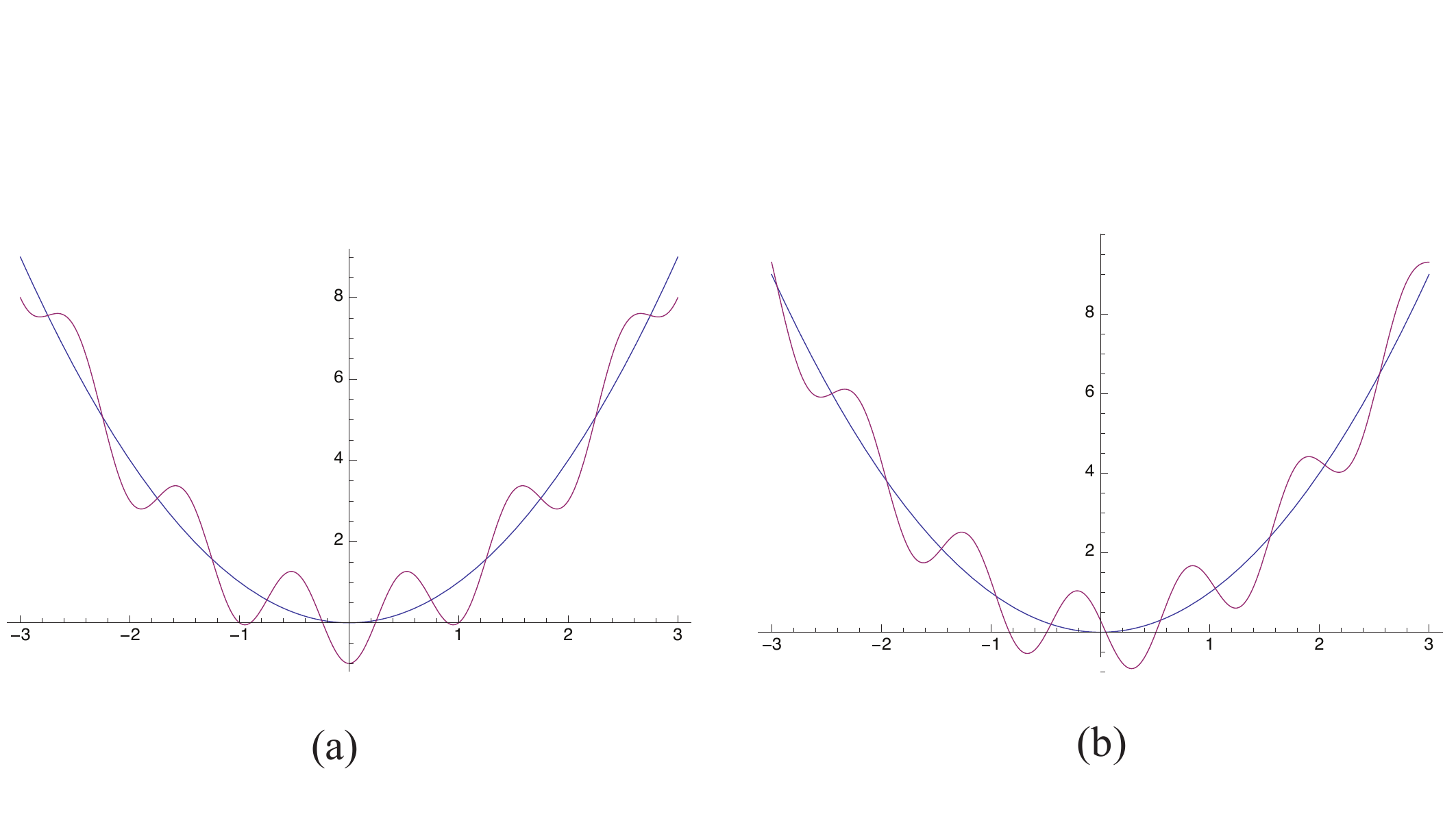}\end{center}
\vspace{-.5cm}
\caption{\label{washboard}The potential energy landscape for the sine-gordon scalar $\phi$, in the strong coupling limit; the pure quadratic potential is superposed on the total potential for reference.  Figure (a) shows the potential energy for $\theta = 0$. Figure (b) shows the potential energy for a positive shift of $\theta$; note that the minimum has shifted to the right.}
\end{figure}


\subsection{Outline}

\S2 describes the perturbative and nonperturbative dynamics of gauge theories coupled to charged fermions. \S2.1\ reviews the classical vacuum structure of the pure gauge theories.  \S2.2\ describes the theories with charged fermions, and their scalar duals.  In \S2.3\ I investigate the interplay between corrections to $E(\theta) \sim \theta^2$ and the onset of instability of a branch; I find that in all of the models discussed the two phenomena occur in the same regime of $\theta$.  In \S2.4\ I discuss the relationship between these theories and the 4d theories discussed above.  \S3\ is an investigation of the 2d sigma model with target space ${\mathbb CP}^N$, in the large N limit.  I extend the calculation of \cite{Witten:1978bc}\ to find the nonlinear corrections to the Maxwell action governing the low-energy dynamics of the ${\mathbb CP}^N$ model. I compute the mass gap as a function of $\theta$, and the interplay between corrections to $E(\theta)$ and the onset of instability for a given branch, and find that the instability becomes relevant before the regime in which corrections to $\CE(\theta) \sim \theta^2$ become important.  \S4\ contains two concluding remarks.

\section{2D gauge theories and $\theta$ angle monodromy}

\subsection{Pure gauge theory}

I will begin with a discussion of the spectrum of pure abelian and non-abelian gauge theory on $S^1$ and on $\RR$.  The spectrum of these theories will map directly to the metastable states of the theories coupled to charged fermions.

\subsubsection{Abelian theory}

Consider a two-dimensional $U(1)$ gauge theory:
\be
	{\cal L} = \frac{1}{4 e^2} F_{\mu\nu}F^{\mu\nu} + \frac{\theta}{4\pi} \epsilon^{\mu\nu}F_{\mu\nu}
\ee
The $\theta$ term is normalized so that the quantization of $F$ ensures that the action shifts by $2\pi\Z$ as $\theta \to \theta + 2\pi$. Here $F_{\mu\nu} = \d_{\left[\mu\right.} A_{\left.\nu\right]}$ is the field strength of an Abelian gauge field; the only nonzero component is $E \equiv F_{01}$.  We will take the $U(1)$ gauge group to be compact.  $\theta$ induces a constant electric field \cite{Coleman:1976uz}, and the energy increases as $\theta^2$.  This is clear in the Hamiltonian formulation of the theory.  Fixing to $A_0 = 0$ gauge, the canonical momentum for $A_1$ is
\be\label{eq:abelian}
	\Pi = \frac{1}{e^2} E + \frac{\theta}{2\pi}
\ee
If the $U(1)$ is compact, then $\Pi = k  \in {\mathbb Z}$.  When space is noncompact, $\Pi$ can be thought of as the charge at infinity (with opposite charge at $-\infty$). There are no local gauge field dynamics, as the gauge freedom $A_1 \to A_1 - \p_1 \Lambda(x^1)$ is unfixed by $A_0 = 0$. $\Pi$ can only change in the presence of charged matter.  The Hamiltonian is
\be\label{eq:abelianham}
	H = \Pi E - {\cal L} = \frac{e^2}{2} \left(\Pi - \frac{\theta}{2\pi}\right)^2
\ee
and is invariant under the shift $\Pi \to \Pi + 1$, $\theta \to \theta + 2\pi$.  For fixed $\theta$, there are a tower of states with energies ${\cal E}_k(\theta) = \frac{e^2}{2}\left(k - \frac{\theta}{2\pi}\right)^2$.  These states reshuffle as one adiabatically increases $\theta$, so that the spectrum is as in Figure \ref{monofig}.  This is the basic phenomenon of "theta angle monodromy".  Although $\theta$ is a periodic variable, the spectrum is only periodic if one simultaneously shifts $\Pi$.  For fixed $\Pi$, one may increase $\theta$ continually, and (\ref{eq:abelianham}) will increase quadratically.  We will refer to this as "large $\theta$". Note that if we promote $\theta$ to a dynamical scalar, the theory is the precise 2D analog of the axion-four form theory studied in \cite{Kaloper:2008fb,Kaloper:2011jz}, as noted in \cite{Seiberg:2010qd}. 

\subsubsection{$SU(N)$, $SU(N)/\Z_N$, and $U(N)$ gauge theories}

Next, consider the theory
\be\label{eq:sunaction}
 	{\cal L} = \frac{1}{4 q^2}\tr \F_{\mu\nu}\F^{\mu\nu}
\ee
Here $\F$ is a nonabelian gauge field strength for $SU(N)$ or $U(N)$. 

When space is noncompact, Witten \cite{Witten:1978ka}\ has shown that this theory has a tower of energy eigenstates, one for every irreducible representation $R$ of $G$, with energies $\CE_R = q^2 C_2(R)$, where $C_2(R)$ is the second Casimir of the representation. These are the analogs of the $\theta$ vacua for the Abelian case, and can be thought of as arising from static charge in the representation $R$ placed at $x = \infty$ together with an antiparticle in the conjugate representation placed at $x = - \infty$. 

When space is an $S^1$, one can show that the configuration space is the configuration space of Wilson lines, up to conjugation by the group.  The states are thus described by the characters $\chi_R(g)$ of the irreducible representations $R$; the Hamiltonian is once again $q^2 C_2(R)$ \cite{Rajeev:1988zb}.\footnote{An alternate quantization, using the gauge connection as the fundamental variable and fixing gauge, leads to an inequivalent spectrum with extra low-lying states \cite{Hetrick:1989vm,Hetrick:1993nq}; we will not address that quantization here.}

The gauge field itself is invariant under actions by the $\Z_N$ center of $SU(N)$, as is any adjoint matter.  If we declare that the true gauge group is $SU(N)/\Z_N$, the theory is labeled by an additional discrete parameter, and for each value of this parameter the spectrum is a restriction of the $SU(N)$ spectrum.  More precisely, if we rotate a given Wilson line $g$ by the center, $g \to \omega g$ with $\omega = e^{2\pi i/N}$, then $\chi_R \to \omega^{N_R} \chi_R$, where $N_R$ is the "N-ality" of the representation $R$ (the number of boxes in the corresponding Young tableaux). The parameter $N_R \mod N$ can be thought of as a discrete $\theta$ term \cite{Chandar:1993ch}.  For a given value of this term, the spectrum of the theory is labeled by representations which share the same $N$-ality.

In the case of $U(N)$, the algebra is that of $SU(N)\times U(1)$. The gauge field strength $F_{\mu\nu} = (dA)_{\mu\nu}$ is an $N\times N$ Hermitian matrix, and we can write the $SU(N)$ piece as:
\be
	{\tilde F}_i{}^j = F_i{}^j - \frac{1}{N} \tr F \delta_i{}^j
\ee
Defining $G = \frac{\tr F}{N} = d B$, eq. (\ref{eq:sunaction}) becomes
\be\label{eq:unaction}
	{\cal L} _{U(N)} = \frac{1}{4 q^2} \tr{\tilde F}_{\mu\nu}{\tilde F}^{\mu\nu} 	
		+ \frac{N}{4 q^2} G_{\mu\nu}G^{\mu\nu}
\ee
Note that if we couple $F$ to fundamental matter, $B$ will couple to this matter with $U(1)$ charge $q$.  Thus, the volume of the $U(1)$ gauge group is $2\pi$.  

Following the prior discussion, the energy eigenstates of the $U(N)$ theory on a circle will take the form
\be
	\psi_{p,R}(\phi,g) = e^{ip\phi} \chi_R(g)
\ee
where $\phi = \oint B$, $p\in \Z$, and $g\in SU(N)$.  As before, different values of $p,R$ correspond to different superselection sectors.  The basic point is that if we write a $U(N)$ matrix $U = e^{i\phi} g$ with $g\in SU(N)$, then the shift $g\to \omega g$ (with $\omega = e^{2\pi i/N}$), $\phi \to \phi - 2\pi/N$ leaves the $U(N)$ matrix is invariant ({\it c.f.} \cite{Guralnik:1997sy}).  The wavefunctions with
\be
	p = N k - N_R + \delta
\ee
will transform as $\psi_{p,R} \to e^{2\pi i \delta/N}$, and correspond to distinct superselection sectors labeled by $\delta$.

In addition, we can add a theta term for the abelian vector field
\be
	L_{\theta} = - \frac{\theta}{2\pi} \tr F_{01} = - N \frac{\theta}{2\pi} G_{01}
\ee
Note the factor of $N$.  We can show that $\theta \equiv \theta + 2\pi$ using either of two arguments.    The first argument (related to that in \cite{Affleck:1984ar}) is that the identification
$(\phi,g) \equiv (\phi - \frac{2\pi}{N}, \omega g)$ means that there are Euclidean configurations on the torus with magnetic flux $\int G_{01} = 1/N$ (attended by $\Z_N$-twisted flux in the $SU(N)$ sector \cite{tHooft:1981sz,Guralnik:1997sy}).  In these cases the theta term shifts by $2\pi$ when $\theta \to \theta + 2\pi$, and the action is invariant.  

The second argument follows from considering the Hamiltonian in the presence of the $\theta$ term,
\be\label{eq:unhamiltonian}
	H = \half q^2 N L \left(\frac{P_{\phi}}{N} - \frac{\theta}{2\pi}\right)^2 + q^2 C_2(R) L
\ee
where $P_{\phi}$ is the momentum conjugate to $\phi$, and $L$ is the circumference of the circle.  Without changing the representation $R$, $P$ can only shift by $N$ without changing the $\Z_{N}$ theta angle.  Thus, the Hamiltonian will be invariant if we shift $P \to P + N k$, $\theta \to \theta + 2\pi k$.

The tower of states for the $U(N)$ theory is richly structured.  Choose $\theta = \delta = 0$.  There is a tower of states with $P = N k$, $H = \half q^2 N k^2 L$.  If $R$ lives in the fundamental, then
$P_{\phi} = N k - 1$, and
\be
	\frac{H}{L} = \half q^2 N \left(k - \frac{1}{N}\right)^2 + q^2 \left(N - \frac{1}{N}\right)
\ee
as $C_2(R_f) = N - \frac{1}{N}$.
If we choose $\delta = 1$ and $R$ the trivial representation, then 
\be
	H = \half q^2 N \left(k + \frac{1}{N}\right)^2 \ .
\ee

\subsection{Gauge fields with charged matter}

In this section we will discuss the spectrum and the semiclassical action for the above gauge theories coupled to charged matter.  The stability of excited states with Abelian or non-Abelian flux will be discussed in \S2.3\ along with the size of quantum corrections to the effective potential $V(\theta)$. 

\subsubsection{Massive Schwinger model}

We begin by reviewing the well-known massive Schwinger model\footnote{A very nice review of the physics of this model and its Sine-Gordon dual can be found in \cite{Kleban:2011cs}.}
\be\label{eq:diractheory}
	{\cal L} = \frac{1}{4 e^2} F_{\mu\nu}F^{\mu\nu} + \frac{\theta}{4\pi} \epsilon^{\mu\nu}F_{\mu\nu}
	+ i {\bar\psi} \left(i \slashed{\d} - \slashed{A} - m\right)\psi\ ,
\ee
and its dual.  The presence of charged fermions renders the excited states of the Abelian vacua metastable; in particular, if one begins in the ground state at $\theta = 0$ and adiabatically increases $\theta$, the system becomes metastable to the pair production of charged fermions \cite{Coleman:1976uz}.  We will discuss this further in \S2.3. 

The massive Schwinger model is dual via bosonization to the Sine-Gordon theory for a single massive scalar \cite{Coleman:1975pw,Coleman:1976uz}.  One first identifies the chiral current as $j_{\mu} = \frac{1}{\sqrt{\pi}} \epsilon_{\mu}{}^{\nu}\partial_{\nu}\phi$. The gauge field coupling to the Dirac fermions becomes, upon integrating by parts, $\sqrt{\pi} \phi F_{01}$, so that $\phi$ couples as an axion.  Upon integrating out $F$, one finds:
\be\label{eq:massivesg}
	{\cal L} = \half (\d\phi)^2 - \half e^2\left(\phi + \frac{\theta}{2\sqrt{\pi}}\right)^2 + c\ m^2\cos 2\sqrt{\pi}\phi
\ee
where $c$ is a constant of order 1.  The potential for $\phi$ is shown in Figure \ref{washboard}, and is a quadratic potential modulated by the cosine term.   As one shifts $\theta$, the minimum of the quadratic terms shifts.   In the limit $e^2 \gg m^2$, the Sine-Gordon theory is weakly coupled, and the cosine term is a small perturbation of the quadratic term in $\phi$.  In this limit, only the vacuum at $\phi + \frac{\theta}{2\sqrt{\pi}} = 0$ is even classically stable.  In this case, the analogy to axion monodromy inflation pertains if we consider $\phi$ as the inflaton; the potential is quadratic modulated by periodic corrections.  In four dimensions, these corrections can lead to interesting results such as resonant non-Gaussianity \cite{Chen:2008wn,Flauger:2009ab,Flauger:2010ja}.

In the limit $m^2 \gg e^2$, there are $\CO(m^2/e^2)$ metastable vacua. As one adiabatically shifts $\theta$, the ground state at $\theta = 0$ becomes metastable for $\theta > \pi$, and remains metastable with increasing $\theta$ until $\theta \sim \CO(m^2/e^2)$  

In the Sine-Gordon theory, $\phi$ represents a fermion-anti fermion bound state or meson (the analog of the $\eta'$ meson in 4d QCD, as it shifts under chiral rotations of the fermions).  The mass of this particle decreases as one adiabatically increases $\theta$, or equivalently as one studies higher and higher-energy metastable states.   To see this, first consider vacua near the ground state, at $\theta = 0$; let $\epsilon = e^2/m^2 \ll 1$. For the $n$th metastable vacuum above the ground state, where $n \eps \ll 1$, the metastable vacuum is at 
$\phi_n = \sqrt{\pi} n (1 - \eps)$; the mass is 
\be
	m_n^2 = V''(\phi_n) \sim e^2 + m^2 \left(1 - 2\pi^2 n^2 \eps^2\right)
\ee
Second, consider the metastable vacua which are close to being classically unstable, again at $\theta = 0$.  These will occur for
\be
	2\sqrt{\pi} \phi_n = 2\pi n + \frac{3\pi}{2} - \delta_n
\ee
Let $2\sqrt{\pi} \phi_{n_{max}} = 2\pi n_{max} + \frac{3\pi}{2} - \Delta$ be the local minimum of (\ref{eq:massivesg}) with highest possible $\phi$. Solving for $V'(\phi_n) = 0$, we find that $0 \leq \Delta \lesssim \sqrt{2\pi\eps}$, and
\be
	\delta_n \sim  \sqrt{\Delta^2 + 2\pi (n_{max} - n)\eps}
\ee
where $n_{max} - n \sim \CO(1)$. The mass of $\phi$ at these vacua are:
\be
	V''(\phi_n) \sim e^2 + m^2 \delta_n
\ee
If $\Delta \lesssim 0$, then the mass of small fluctuations of $\phi$ about $\phi_n$ (drops from $m^2$ to $\CO(m e)$ for the highly metastable vacua. If $\Delta \sim \sqrt{\eps}$, the meson mass drops to $\CO(m^{3/2} e^{1/2})$.

\subsubsection{The gauged massive Thirring model}

Next, we add the operator
\be\label{eq:fourf}
	\delta L = - \half g \bar{\psi}\gamma_{\mu}\psi \bar{\psi}\gamma^{\mu}\psi
\ee
to the massive Schwinger model.   When $e^2 \to 0$, the theory is known as the massive Thirring or massive Luttinger model, and
it is dual to the Sine-Gordon model for a scalar living on a circle with (dimensionless) radius $R(g) = \sqrt{\frac{1 + g/\pi}{4\pi}}$:\footnote{This is related to \cite{Coleman:1974bu}\ by the redefinition $\phi \to \phi R$, with $R$ denoted $\beta$ in that text.}
\be\label{eq:SGperturbed}
	L = \half R^2 (\p\phi)^2 - R^2 \mu^2 \cos \phi\ 
\ee
The map between $\mu$ and $m$ depends on the renormalization scheme on each side of the duality \cite{Coleman:1974bu,Mandelstam:1975hb}.  Let $\Lambda_{IR}$ define the "normal-ordering scale".\footnote{$\Lambda_{IR}$ is  by decomposing a scalar field in creation and annihilation operators corresponding to a scalar of mass $\Lambda_{IR}$, and normal ordering with respect to those operators. Similarly, one can define the scalar propagator as $\langle \phi(x)\phi(0)\rangle = - \frac{1}{4\pi} \ln \Lambda_{IR}^2 x^2$, and compute correlation functions of $e^{i\beta\phi}$ using Wick's theorem with this propagator.} The relation between fermionic and bosonic mass parameters is then $m \Lambda_{IR} = \mu^2 R^2$. Near $\phi = 0$, the mass of the canonically normalized scalar $R\phi$ is $\mu^2$, so that $\mu$ is the most natural candidate for $\Lambda_{IR}$\footnote{A different choice would lead to finite renormalization effects in the definition of the composite operator $\cos\phi$; this is nicely explained in \cite{Coleman:1974bu}.}; in this case, $\mu = m/R^2$.  

When $\mu^2 > 0$, $R$ changes the RG flow of the theory \cite{Kosterlitz:1974sm,Coleman:1974bu,Amit:1979ab}. For $R > 1/\sqrt{8\pi}$, the cosine term is relevant (this includes $R = 1/\sqrt{4\pi}$, $g = 0$); the theory flows to large $R$, $m^2$ \cite{Amit:1979ab}. For $R < 1/\sqrt{8\pi}$ the cosine term is irrelevant and the theory is nonrenormalizable; the theory flows to a free theory with $R \leq 1/\sqrt{8\pi}$.  

Note that at large $R$, when the theory (\ref{eq:SGperturbed}) is semiclassical, increasing $R$ increases the mass of the Sine-Gordon kinks {\it at fixed} $\mu$.  This is because the overall action (\ref{eq:SGperturbed}) increases as $R^2$, although the equation of motion is independent of $R$.  The resulting mass of a kink thus scales as 
\be
	m_{kink} \sim R^2 \mu = \frac{1}{4\pi}\left(1 + \frac{g}{\pi}\right) \mu
\ee
For $\Lambda_{IR} = \mu$, this is still the mass parameter $m$ appearing in the fermion action (\ref{eq:fourf}).  

Now consider adding (\ref{eq:fourf}) to the Schwinger model (\ref{eq:diractheory}).  The scalar dual is:
\be
	L = \half R^2 (\p\phi)^2 + R^2 \mu^2 \cos \phi - \left(\frac{\phi + \theta}{4\pi}\right) \epsilon^{\mu\nu}F_{\mu\nu} - \frac{1}{4e^2} F_{\mu\nu}F^{\mu\nu}
\ee
The coefficient of the $\phi-F$ coupling is set by the fact that $\phi$, like $\theta$, has periodicity $2\pi$. Rescaling the gauge field and gauge coupling $F = R^2 \tF$, $e^2 = \te^2 R^2$, the action becomes:
\be\label{eq:rescaledsg}
	L = R^2 \left[ \half (\p\phi)^2 + \mu^2 \cos\phi - \frac{(\phi + \theta)}{2\pi} \tF_{01} -\frac{1}{2\te^2} \tF_{01}^2\right]
\ee
$R$ appears as a loop-counting parameter. 

If we fix the canonical momentum of $F$ to vanish, then we can write the Hamiltonian as:
\be
	H = \frac{1}{2R^2} \Pi_{\phi}^2 + R^2 \left( - \mu^2\cos\phi + \half\te^2 (\phi + \theta)^2 \right)
\ee
In the limit $\te^2 < \mu^2$ or $e^2 < m^2/R^2$, the theory has a global minimum and $\CO(\te^2/\mu^2)$ metastable vacua.  In this case, the scalar mass close to the minimum is still $\mu^2$, and it makes sense to continue to set $\Lambda_{IR} = \mu$.

In the limit $\te^2 > \mu^2$, the quadratic term dominates and there are no metastable vacua.  In this latter case, the physical mass of $\phi$ is $\te^2 = e^2/R^2$. This relation is the 2d version of the relation found in \cite{Kaloper:2011jz}\ between the axion mass, axion decay constant, and unit of four-form flux quantization. Note that in this case, the natural value of $\Lambda_{IR}$ is the physical mass $\te$ of the canonically normalized scalar ({\it c.f.}\ \cite{Affleck:1985wa}).  Adopting this, we find that $\mu^2 R^2 = m \te$.

\subsubsection{The $U(N)$ 't Hooft model}

Now consider the theory
\be\label{eq:unthooft}
	L = \frac{1}{4 e^2} \tr F_{\mu\nu}F^{\mu\nu} + \frac{\theta}{4\pi} \tr \epsilon^{\mu\nu}F_{\mu\nu}
	+ {\bar\psi}^i \left(i \delta_i{}^j \slashed{\d} - \slashed{A}_i{}^j - \delta_i{}^jm\right)\psi_j\
\ee
Here $F = dA + A^2$ is the field strength for gauge group $G = SU(N)$, or $U(N)$.  We take the quarks $\psi$ to transform in the fundamental representation $R_f$ of $G$. The $U(1)$ charge corresponds to $N$ times the baryon number.  

In the case $G = SU(N)$, the excited states labeled by representations $R$ are all metastable \cite{Witten:1978ka}; if $R\otimes R_f$ contains a representation $R'$ such that $C_2(R') < C_2(R)$, the system can make a transition from the state labeled by $R$ to the state labeled by $R'$, via pair production of  quarks.

When $G = U(N)$ the Abelian flux leads to a richer story.  The Lagrangian is:
\be\label{eq:uncharged}
	L = \frac{1}{4 q^2} \tr \tF_{\mu\nu}\tF^{\mu\nu} + \frac{N}{4 q^2} G_{\mu\nu}G^{\mu\nu}
	+ \frac{N \theta}{2\pi} G_{01}
	+ {\bar\psi}^i \left( i \delta_i{}^j \slashed{\d}  - \slashed{\tA}_i{}^j - \delta_i{}^j \slashed{B} - \delta_i{}^j m\right)\psi_j\
\ee
where $\tF = d\tA + \tA^2$, $G = dB$ are the fields that appear in (\ref{eq:unaction}).  As discussed in \S2.1.2, absent the fermions the theory has a rich landscape of states labeled by $\theta, R$. Many of these are rendered metastable or unstable by the inclusion of charged matter.  Depending on the value of $\theta$, $R$ one can pair produce quarks which carry both $SU(N)$ and $U(1)$ charge, or baryons which have $U(1)$ charge alone. We will find in \S2.3\ that the dominant decay channel is via baryon pair production.

We will focus on the weakly-coupled limit $e^2 \ll m^2$. (We will see presently that it is $e^2/m^2$ which governs the stability of the theory against baryon production; $m \sim e^2 N$ is the threshhold at metastable vacua exist for which quark pair production is energetically favorable). For discussions of the strong coupling dynamics when either the $U(1)$ or $SU(N)$ symmetry is gauged, see for example \cite{Steinhardt:1980ry,Affleck:1985wa,Frishman:1992mr,Abdalla:1995dm}.  Near the true vacuum of the theory, this limit is best studied via the fermionic presentation.  However, far along a metastable branch parametrized by $\theta$, the bosonic dual is a useful presentation of the theory. 

The nonabelian bosonization of  (\ref{eq:uncharged}) follows \cite{Coleman:1974bu,Witten:1983ar,Affleck:1985wa}.  The bosonic degrees of freedom consist of a scalar $\phi$ with radius $\sqrt{\pi N}$, and an $SU(N)$ matrix $g$, with Lagrangian:
\begin{eqnarray}
	S & = & S_{WZW}(g,\tA) + \int d^2 x \left[ \half (\p\phi)^2 - \left(\frac{\sqrt{N}}{2 \sqrt{\pi}}\phi  + \frac{N \theta}{2\pi}\right) G_{01} - \frac{N}{2 e^2} G_{01}^2 - \frac{1}{2 e^2} \tr \tF_{01}^2 \right.\nonumber\\
	& & \ \ \ \ \ \left. + \mu^2
		\left(\tr g e^{i \sqrt{\frac{4\pi}{N}} \phi} + c.c.\right)_{\Lambda_{IR}}\right]
		\label{eq:naboson}
\end{eqnarray}
Here $\Lambda_{IR}$ is the mass scale at which we normal order the composite operator $\left(\tr g e^{i \sqrt{\frac{4\pi}{N}} \phi}\right)$, and $\mu^2 = m \Lambda_{IR}$. $S_{WZW}$ is the gauged $SU(N)$ WZW action at level $k = 1$ \cite{Affleck:1985wa}.  

The interactions between the $SU(N)$ and $U(1)$ bosons clearly depend only on the eigenvalues of $g$, and for $\phi = 0$ the potential energy is minimized by $g = {\bf 1}$.  To get a handle on the large-$\phi$ limit,  we will first study the Abelian bosonization of the theory following \cite{Baluni:1980bw,Steinhardt:1980ry} (see \cite{Frishman:1992mr,Abdalla:1995dm}\ for a further review and references). We set
\be
	j_{\mu}^i = {\bar\psi}^i \gamma_{\mu} \psi_i = \frac{1}{\sqrt{\pi}} \epsilon_{\mu}{}^{\nu} \p_{\nu}
		\phi^i
\ee
Following \cite{Baluni:1980bw,Steinhardt:1980ry}, we write 
\be
	\phi^i = \frac{\phi}{\sqrt{N}} + \sum_j M_{N-i,N-j} \chi_j\ ; i \in (1,\ldots N),\ j\in(1,\ldots,N-1)
\ee
where $\phi$ is the same scalar as in (\ref{eq:naboson}) and $\chi_j$ couples only to the $SU(N)$ gauge field. The $N\times (N-1)$ matrix $M$ is defined as
\be
	M_{N-i,N-j} = \left\{ \begin{array}{ll} 0 & j < i -1 \\ - \sqrt{\frac{j}{j + 1}} & j = i -1 \\
		\frac{1}{\sqrt{j(j+1)}} & j > i -1 \end{array} \right.\ 
\ee
It is easy to show that $\sum_i M_{N-i,N-j} = 0$, $\sum_i M_{N-i,N-j} M_{N-i,N-k} = \delta_{j,k}$. If we move to the Hamiltonian form of the theory and integrate out the gauge fields after an appropriate gauge fixing, we find \cite{Baluni:1980bw,Steinhardt:1980ry}:
\begin{eqnarray}
	H & = & \half \pi_{\phi}^2 + \half (\p_1\phi)^2 + \sum_{i = 1}^{N-1} \left(\half \pi_i^2 + \half(\p_1\chi^i)\right) + V(\phi,\chi^i)\nonumber\\
	V & = & \frac{e^2}{2\pi} \sum_i (\chi^i)^2 + \frac{e^2}{2\pi} \left(\phi + \sqrt{\frac{N}{\pi}}\theta\right)^2 \nonumber\\
	& & \ \ \ \ - m\Lambda_{IR} \cos\left(2\sqrt{\frac{\pi}{N}}\phi\right)\sum_i \cos\left(\sum_j M_{N-i,N-j}\chi_j\right)
	\nonumber\\
	& & 	\ \ \ \ + m\Lambda_{IR} \sin\left(2\sqrt{\frac{\pi}{N}}\phi\right)\sum_i \sin\left(\sum_j M_{N-i,N-j}\chi_j\right)\nonumber\\
	& & \ \ \ \ - \Lambda^2 \sum_{i,j\neq i} \frac{\sin(\phi^i - \phi^j)}{\phi^i - \phi^j}\label{eq:evals}
\end{eqnarray}
where $\Lambda_{IR}$ is the normal-ordering scale.  At large $m^2$, we choose $\Lambda = \frac{m}{4\pi}$, which we will find corresponds (near the true vacuum) to bosons of mass $m^2$. (This is in distinction to strong coupling, $e^2 \gg m^2$, which is the focus of study in \cite{Baluni:1980bw,Steinhardt:1980ry,Affleck:1985wa}.)  At strong coupling, the final term in $V$ is clearly difficult to normal-order and is best represented merely as a complicated nonlinear function of $\phi^i - \phi^j$ \cite{Steinhardt:1980ry}.  However, when $m^2 \gg e^2$, this term has a clear minimum at $\phi^ i = \phi^j$, where $\chi = 0$, and we can define the potential by a power series expansion about this point:
\be
	\sum_{i,j\neq i} \frac{\sin(\phi^i - \phi^j)}{\phi^i - \phi^j} \sim 1 - \frac{1}{6} \sum_{i,j\neq i} (\phi^i - \phi^j)^2 	\sim 1 - \frac{N}{3}\sum_k \chi_k^2
\ee
It is easy to see that $\chi^i = 0$ remains a classically stable solution to the equations of motion as $\phi$ increases.  The third term in (\ref{eq:evals}) will contribute zero to the mass of $\chi$ at this point.  As we adiabatically increase $\theta$ so that $\phi$ is pushed to some highly metastable branch, $\cos\left(2 \sqrt{\frac{\pi}{N}}\phi\right)$ at the metastable branch will become small.  The final term in (\ref{eq:evals}) will still be of order $m^2$ and will continue to dominate.  We can thus integrate out $\chi_i$ to find the effective dynamics for $\phi$:
\be
	V(\phi) = \frac{e^2}{2\pi} \left(\phi + \sqrt{\frac{N}{\pi}} \theta\right)^2 
		- \frac{m^2 N}{4\pi} \cos\left(2 \sqrt{\frac{\pi}{N}}\phi\right)
\ee
The result is equivalent to (\ref{eq:naboson}) if we set $g = {\bf 1}$.  Rescaling $\phi = \sqrt{N} \varphi$, we find:
\be
	S = N \int d^2 x \left[ \half (\p\varphi)^2 - \left(\frac{2 \sqrt{\pi}\varphi - \theta}{2\pi}\right) G_{01}
		- \frac{1}{2e^2} G_{01}^2 - \frac{m^2}{4\pi} \cos 2\sqrt{\pi}\varphi\right]\label{eq:largenscalaract}
\ee
As we increase $\theta$, the theory becomes unstable at precisely the same value as it does in the theory (\ref{eq:diractheory}).  In this case, however, the kink solitons which get pair-produced are baryons \cite{Steinhardt:1980ry} with $U(1)$ charge $N$ and mass $N m$.

The action (\ref{eq:unthooft}) was discussed in \cite{Affleck:1985wa}\ when either only the $U(1)$ or $SU(N)$ gauge symmetries were gauged, in the limit that $e^2/N \gg m^2$ for the $U(1)$ theory, or $e^2 \gg m^2$ for the $SU(N)$ theory.  In these cases, one naturally normal orders the theory at the scale of the gauge coupling $e^2/N$ or $e^2$. For the $U(1)$ case, the gauging gives a mass to the boson $\phi$ which overwhelms the potential energy dual to the fermion mass, and there are no metastable states.  For the $SU(N)$ case, the gauging leads to a mass for $g$ and an expectation value $\langle\tr g \rangle \sim N$ and the low energy effective theory is a Sine-Gordon model.  We refer the reader to \cite{Affleck:1985wa}\ for a more complete discussion.

%
%

\subsubsection{The 't Hooft model with U(1) current-current interactions}

As in \S2.2.1, we will add a four-fermion term equal to the square of the $U(1)$ current:
\be\label{eq:thooftfourf}
	\delta L = - \half g {\bar\psi}^i\gamma^{\mu}\psi_i{\bar\psi}^j\gamma_{\mu}\psi_j
	= \frac{N g}{2\pi} (\p\phi)^2
\ee
The resulting Lagrangian for $\phi$ is:
\be
	S = N \int d^2 x \left[ \half \tR^2 (\p\varphi)^2 - \left(\frac{\sqrt{\pi}\varphi - \theta}{2\pi}\right) G_{01} 
		- \frac{1}{2e^2} G_{01}^2 - \frac{m^2\tR^2}{4\pi} \cos 2\sqrt{\pi}\varphi\right]
\ee
where 
\be
	\tR^2 = \frac{1}{4\pi} \left(1 + \frac{Ng}{\pi}\right)
\ee
The cosine term in multiplied by a factor of $\tR^2$ so that the physical mass of $\phi$ does not change.  As in \S2.2.1, we can rescale $G = \tR^2 \tG^2$, $e^2 = \tR^2 \te^2$, and find
\be
	S = N\tR^2 \int d^2 x \left[ \half (\p\varphi)^2 - \left(\frac{\sqrt{\pi}\varphi - \theta}{2\pi}\right) \tG_{01} 
		- \frac{1}{2\te^2} \tG_{01}^2 - \frac{m^2}{4\pi} \cos 2\sqrt{\pi}\varphi\right]\label{eq:lnintscalaract}
\ee
Again, begin with the system in the true ground state at $\theta = 0$, and begin increasing $\theta$, staying on a given branch of the monodromy potential as it becomes metastable.  Let $\theta(m,e)\sim m^2/e^2$ be the value of $\theta$ for which that branch becomes unstable in theory (\ref{eq:largenscalaract}).  In the theory (\ref{eq:lnintscalaract}), a given metastable branch becomes unstable at the same value of $\theta = \theta(m,\te)\sim m^2/\te^2$.  At large $\tR$, $\te \ll e$ and the range of $\theta$ is extended.

\subsection{Stability and quantum corrections}

The motivation for this work was the study of four-dimensional "axion monodromy" models in which the $\theta$ term becomes a dynamical axion. Models which are at all calculable appear to lead to a potential which starts quadratically in the axion near the bottom of the potential, and then flattens out, running as a power $\phi^p$ with $p < 2$ at large $p$\cite{Silverstein:2008sg,McAllister:2008hb,Dong:2010in}, or even going as $V_0(1 - (\mu/\phi)^n)$ \cite{Dubovsky:2011tu}. We would like to know if this occurs in two dimensions. On the other hand, we will find that the probability for the metastable vacuum to decay also increases for larger $\theta$, as also occurs in \cite{Dubovsky:2011tu}.  In that work, there is a range along a given metastable branch where the potential is flat and transitions to lower-energy vacua are suppressed.   In this section we will argue that for all of the models studied in \S2.2, the instability kicks in as soon as the nonquadratic corrections are $\CO(1)$.

\subsubsection{The Schwinger model}

For $\Pi = 0$ and $|\theta| > \pi$, $E = \frac{e^2\theta}{2\pi}$, the branch starting at the ground state of $\theta = 0$ becomes metastable, and the theory becomes unstable to pair production of the charged fermions \cite{Coleman:1976uz}.  This has the effect of shifting $\Pi \to \Pi - 1$ between the charges.  

For $\theta \ll m^2/e^2$, the decay probability can be described by a Euclidean worldline instanton; a circular trajectory for the charged particle surrounding a region of electric field with strength $E - e^2$.  The action is \cite{Brown:1987dd,Brown:1988kg}:
\be\label{eq:bubbleaction}
	S_{inst} = \frac{\pi m^2}{\Delta {\cal E}} = \frac{\pi m^2}{E - \half e^2}
\ee
where $\CE$ is the difference between the energy densities inside and outside of the bubble. 
One might expect that the states remain metastable so long as the action $S < \CO(1)$, that is
\be
	E - \half e^2 = e^2\left(\frac{\theta}{2\pi} - \half\right) < \pi m^2
\ee
For the Schwinger model this is in fact that case, as we will see by summing up the instantons and by studying the Sine-Gordon dual.  Indeed, $m^2/e^2$ is the only dimensionful ratio in the theory.

In the limit $m^2 \gg e^2$, the fermionic theory is weakly coupled and the loop expansion should be a good one.  At one loop, the effective Lagrangian can be easily calculated after \cite{Schwinger:1951nm} (see also chapter 4.3 of \cite{Itzykson:1980rh}):
\be\label{eq:fermeffact}
	L_{eff} = - \frac{i}{4\pi} \int_{\eps}^{\infty} \frac{ds}{s} E \coth (Es) e^{- i m^2 s}\ ,
\ee
where $\eps$ is the proper time cutoff. The imaginary part of the effective Lagrangian is the decay rate per unit time per unit length:
\be
	\Gamma = - \frac{E}{4\pi} \sum_{n=1}^{\infty} \frac{1}{n} e^{-n \pi m^2/|E|} = 
		- \frac{E}{4\pi} \ln \left(1 - e^{-\pi m^2/|E|}\right)
\ee
For $|E| \ll m^2$ this is clearly a sum over multiple instantons.  For $|E| \gg m^2$ $\Gamma \sim E \ln \frac{|E|}{m^2}$ diverges logarithmically in E.  

We can also study this decay process in the Sine-Gordon theory, where the potential for $\phi$ is illustrated in Figure \ref{washboard}.  As one increases $\theta$ adiabatically the cosine modulation shifts.  As $\theta$ is dialed past $\theta = \pi$, the ground state evolves to the lowest-lying metastable state.  As $\theta$ continues to increase, the energy of this state gets higher and higher; as $\theta$ increases above $2\pi (n - \half)$, there are $n-1$ lower-energy metastable states and the ground state with lower energy.   For such metastable states, instability proceeds via pair production of "kink" solitons interpolating between neighboring metastable vacua.  For $\theta \sim\ (few)\times 2\pi$, the probability can be computed in the bosonic picture using the "thin wall" approximation \cite{Kleban:2011cs}.   For false vacua at energies close to the region of classical instability, the thin wall approximation breaks down; the barrier height gets low, and the separation between adjacent vacua becomes small.  In this regime, we expect semiclassical techniques to fail. Furthermore, the potential energy of the classical minima of the false vacua will begin to be larger than that of the top of the barrier separating the next two lower energy minima.  It is then possible for the system between the kinks to overshoot that barrier and continue to evolve.

The next question is whether the overall quadratic envelope of $E(\theta)$  might begin to steepen or flatten when $\theta$ is large.  First, recall that before including fermion loops, we can integrate out the gauge field and find the energy as a function of $\theta$ to be
\be\label{eq:quadraticpot}
	\CE_{class} = \frac{e^2}{2} \left(\frac{\theta}{2\pi}\right)^2\ .
\ee
Fermion loops will induce, at low energies, corrections of the form 
\be\label{eq:nonlincorr}
	\Delta \CL \sim \sum_k c_k \frac{\tr F^k}{m^{2k-2}}
\ee
for small $E^2/m^2$, where $m^2$ is the fermion mass, and $c_k$ are some dimensionless coefficients which can be computed as a power series in $e^2/m^2$. (Super-renormalizability implies that we need not worry about the cutoff dependence). The coefficients $c_n$ can be computed exactly at the one-loop level (\cf\ \cite{Blau:1988iz}). Eq. (\ref{eq:nonlincorr}) includes a renormalization of the gauge kinetic term, shifting the coupling by $e^2 \to \te^2 = e^2/(1 - e^2/(6\pi m^2))$; in the weak coupling limit this is a small shift.  In the Abelian case, integrating out the gauge fields leads to a modification of the quadratic potential $\CE(\theta) \sim e^2 \theta^2$, to one of the form
\be
	\CE(\theta) = e^2 \left(\frac{\theta}{2\pi}\right)^2 \sum_k \left(\frac{e^2 \theta}{m^2}\right)^k
\ee
The upshot is that the $k \geq 1$ corrections become important precisely as the theory becomes classically unstable.  

Nonetheless, let us compute the leading $\CO(\theta^4)$ correction to $\CE(\theta)$.  In the limit $E \ll m^2$, we can expand (\ref{eq:fermeffact}) out to quartic order in $E$.  Ignoring the leading quadratic divergence (which renormalizes the cosmological constant), the combined tree-level and leading one-loop terms in the effective action are:
\be
	L = \frac{E^2}{2\te^2} - \frac{1}{90\pi} \frac{E^4}{m^6} - \frac{\theta}{2\pi} E
\ee
The canonical momentum is:
\be
	\Pi = \frac{1}{\te^2} E - \frac{2}{45\pi} \frac{E^3}{m^6} - \frac{\theta}{2\pi}
\ee
The corresponding Hamiltonian is:
\be
	H = \frac{1}{2\te^2} E^2 - \frac{1}{30\pi} \frac{E^4}{m^6}
\ee
This appears to flatten as a function of $E$.  However, if we study the $\Pi = 0$ branch, for which
\be
	E \sim \te^2 \theta + \frac{2 \te^8}{45\pi m^6} \left(\frac{\theta}{2\pi}\right)^3\ ,
\ee
we find that
\be
	\CE(\theta) = \frac{\te^2}{2} \theta^2 + \frac{\te^8}{90\pi m^6} \left(\frac{\theta}{2\pi}\right)^4 + \ldots
\ee
Thus, the effect of the fermion loops is to slightly steepen $\CE(\theta)$.

When $E \gtrsim m^2$, the theory is clearly unstable. Still, one can compute the real part of (\ref{eq:fermeffact}) in this regime.  Defining $t = m^2 s$, and expanding the hyperbolic cotangent in a power series in $e^{- 2 E t/m^2}$, one finds that the leading $E$-dependent term is:\footnote{This is consistent with the Lorentzian continuation of \cite{Blau:1988iz}.}
\be
	L_{eff} \sim - \frac{E}{4}\ ,
\ee
which gives a finite renormalization of the $\theta$-term.  $L_{eff}$ is clearly subleading to the classical action $E^2/(2 e^2)$; the one-loop term does not dominate even at large $E$.  

In the limit $m^2 \gg e^2$, the cosine potential provides a small modulation of the quadratic potential.  For fixed $\theta$, $\phi$ will always slide down to minimize the potential energy term in (\ref{eq:massivesg}).  One could consider $\phi$ as the axion; this model is then the 2d version of that studied in \cite{Kaloper:2011jz}.  As in that work, periodicity in $\phi$ prevents any direct corrections of the form $\phi^n$; corrections to the effective potential for $\phi$ will arise from corrections of the form (\ref{eq:nonlincorr}).  There is no reason for the small cosine term to significantly flatten the potential.  For vanishing $m^2$, the Sine-Gordon theory is Gaussian, with a linear coupling to the Abelian gauge field, and the cosine perturbation does not grow at large distances relative to the tree-level scalar potential, unlike many marginal perturbations in field theory.

Thus, if the function $E(\theta)$ deviates from a quadratic potential along a metastable branch before the branch becomes classically unstable, the deviation will have to arise from a coupling of the gauge field to additional lighter degrees of freedom which are not also charged to that gauged field.  An obvious possibility is to let the gauge coupling $e^2$ depend on a scalar $\psi$ with mass ${\tilde m} \ll m$.  Related couplings were discussed various four-dimensional models in \cite{Dong:2010in,Kaloper:2011jz}.

\subsubsection{The gauged Thirring model}

Next, we consider the Schwinger model plus the term (\ref{eq:fourf}), in the limit $g \gg 1$.  In this limit, even for $e^2 \ll m^2$, perturbation theory for the fermonic theory fails.  However, we will see that the bosonic dual (\ref{eq:rescaledsg}), with $R^2 = (1 + g^2/\pi)/4\pi$, is clearly semiclassical in this limit, so the theory can be put under control.

Again, as we adiabatically increase $\theta$, the ground state flows to a metastable state.  For $\theta \sim (few)\times2\pi$, the instability occurs through the pair production of scalar kinks.  We expect the instanton action to scale as $R^2$. It is clear that the mass of a scalar kink is $m_{kink} \sim R^2 \mu$.  The energy difference between vacua if one shifts $\theta \to \theta - 2\pi$ is $\Delta \CE \sim R^2 (\tE - \half \te^2)$, where $\tE = \tF_{01}$.  Thus, we expect the instanton controlling the pair production of kinks and antikinks to have the action
\be
	S_{inst,R} = \frac{\pi m_{kink}^2}{\Delta \CE} \sim \frac{\pi R^2 \mu^2}{\tE - \half \te^2}
\ee
One might expect that the theory becomes unstable when $\tE \gtrsim R^2 \mu^2$.  However, this argument is misleading.  As with the Schwinger model, the thin wall approximation will start to break down for instantons mediating the decay of sufficiently high-energy metastable states.  It is clear from (\ref{eq:rescaledsg}) that the theory becomes unstable when $\tE \sim\te^2 \theta \sim \mu^2 \ll R^2\mu^2$.  The point is that the action takes the form $R^2 f(\te^2\theta/\mu^2)$ where $f$ is getting small as $\te^2\theta/\mu^2 \to 1$

In the limit $R^2 \to \infty$, for $\tE < \mu^2$, the corrections to the effective action for $F_{01}$ can be found semiclassically by solving the equation
\be
	\mu^2 \sin\phi = - \frac{\tF_{01}}{2\pi}
\ee
It is clear that $\phi$ is a function of $\frac{\tE}{\mu^2}$, and that this ratio controls corrections to the leading action for $\tF$.  The corrections become important precisely when the theory becomes completely unstable.

\subsubsection{The $U(N)$ 't Hooft model}

Next we turn to (\ref{eq:unthooft}), in the limit  $m^2 \gg q^2$, $N \gg 1$.   Begin with the vacuum $\theta = 0$, $R = ({\rm trivial})$, and adiabatically increase $\theta$.  Two kinds of particles can be pair produced. A quark-antiquark pair will change the representation to $R_f$ and will change $P_{\phi}$ in (\ref{eq:unhamiltonian}) by one unit. One may also pair-produce baryons, bound states of $N$ quarks with $U(1)$ charge $N$ and vanishing nonabelian charge.  The resulting state will have vanishing nonabelian flux, and will have $P_{\phi}$ shifted by $N$.  This is physically equivalent to keeping $P_{\phi}$ vanishing and shifting $\theta$ by $2\pi$.  

As $\theta$ is increased, baryon pair production is allowed before quark pair production.  The former will occur as soon as $\theta > \pi$.  The baryons have mass $c N m$ with $c$ an order 1 constant.\footnote{Check: I am assuming that I can use semiclassical reasoning, with the bare parameters, in this limit.}  The semiclassical action will be:
\be
	S = \frac{\pi m_{baryon}^2}{\Delta E} = \frac{\pi N \mu^2}{q^2 \left(\frac{\theta}{2\pi} - \half\right)}
\ee
Quark pair production will be allowed when 
\be
	\half q^2 N \left(\frac{\theta}{2\pi}\right)^2 > \half q^2 N \left(\frac{\theta}{2\pi} - \frac{1}{N}\right)^2 + C_2(R_f)
\ee
Since $C_2(R_f) = N - \frac{1}{N}$, this condition means that
\be
	\Delta E = q^2 \left( \frac{\theta}{2\pi} - N + \frac{1}{2N}\right) > 0
\ee
Thus, confinement screens the theory against quark pair production for $\theta/2\pi < N - 1/2N$.

In the large-N limit, the dual Sine-Gordon theory (\ref{eq:largenscalaract}) becomes semiclassical.  The analysis of stability and quantum corrections for this action is identical to that in \S2.3.2. The theory becomes unstable to baryon condensation when $\theta > m^2/e^2$.  This occurs above the threshold for quark-anti quark pair production if $m^2 \gg N e^2 \equiv \lambda$, with $\lambda$ the 2d 't Hooft coupling.

\S2.1.2\ described additional states with nonabelian flux which become metastable in the presence of dynamical quarks.  For example, consider the state with $\theta = \Pi = 0$, but with nonabelian flux in the fundamental representation.  (This could arise from placing quarks at infinity, and canceling the abelian flux by shifting the $\theta$ term).  The energy density of this state is $E_N = q^2 \left(N - \frac{1}{N}\right)$.  Pair production of quarks screens this flux, and generates $U(1)$ flux, shifting $P_{\phi}$ in (\ref{eq:unhamiltonian}) by one unit.  The final energy will be $E = \frac{q^2}{2N}$

\subsubsection{The 't Hooft model with $U(1)$ current-current interactions}

One again, when $Ng$ in (\ref{eq:thooftfourf}) is large enough that $N\tR^2 \gg 1$, the scalar theory (\ref{eq:lnintscalaract}) becomes semiclassical.  The only difference from the previous two sections is the functional form of the prefactor.  Once again, the theory becomes classically unstable precisely at the values of $\tG_{01}$ that the quartic and higher terms in the effective action for $\tG$ kick in.

\subsection{Relation to four dimensional models}

In this section we would like to compare the theories studied here to a closely analogous four-dimensional model of axion monodromy \cite{Kaloper:2011jz}.  The remainder of the paper will not depend on this section.

In \cite{Kaloper:2011jz}\ the authors studied a four-dimensional model:
\be\label{eq:fourform}
	S = \int d^4 x \sqrt{g} \left( m_{pl}^2 R^{(4)} - \frac{1}{48} (F^{(4)})^2 + \half (\d\phi)^2 - \mu \phi F^{(4)} \right)
\ee
closely related to the theories studied here. Here $F^{(4)}$ is a 4-form field strength for a 3-form potential; compactness of the associated gauge group ensures that $F$ is quantized as $F^{(4)} = q n \epsilon^{(4)}$, where $q$ has dimension 2, and $\epsilon^{(4)}$ is the 4d volume form.  $n$ can only change via membrane nucleation. $R$ is the Ricci scalar for the metric $g$, and $m_{pl}$ is the 4d reduced Planck mass. $\phi$ a 4d pseudoscalar with field space periodicity $f_{\phi}$. $\mu$ is a mass parameter. The theory is periodic under shifts $\phi \to \phi + f_{\phi}$, $n \to n - 1$ so long as 
\be\label{eq:percond}
	q = \mu f_{\phi}\ .
\ee
In \cite{Kaloper:2011jz}, the scalar field $\phi$ was the inflaton, and the authors considered monodromy in this variable.  One could also consider terms sinusoidal in $\phi$, but for slow-roll inflation to work these terms must be suppressed.  At fixed $\phi$, the theory has a set of metastable configurations labelled by an integer $n$, with energy $(q n - m \phi)^2$.   For fixed $n$, $\phi$ is a massive scalar field. If $m$ is of order $10^{13}\ GeV$ this leads to a viable model of inflation, so long as membrane nucleation is suppressed, and light moduli do not couple too strongly to $F,\phi$.  

The action for membrane nucleation is
 \be
	S = \frac{27 \pi^2}{2} \frac{\sigma^4}{(\Delta V)^3}
\ee
where $\sigma \equiv M_T^3$ is the membrane tension, and ${\Delta V}$ is the difference in potential energy density between the exterior and interior of the bubble At tree level, the equations of motion for $F^{(4)}$ give $V = \frac{1}{48} F^2$; with $F_{0123} = q n$, we can write $\Delta V = q F_{0123} \equiv {\tilde F}_{0123}$, and then:
\be
	S = \frac{27 \pi^2}{2} \frac{\sigma^4}{(\tilde F)_{0123}^3}
 \ee
Here ${\tilde F}$ has kinetic term $\frac{1}{48 q^2} {\tilde F}^2$; this normalization is closer in spirit to that we have chosen for the 2d Maxwell field.

Ref. \cite{Kaloper:2011jz}\ studied quantum corrections to the tree-level dynamics. By itself, quantum corrections generated by loops of $\phi$ and of the graviton in (\ref{eq:fourform}) do not spoil inflation \cite{Linde:1987yb}. The crucial question is whether additional degrees of freedom (as any UV-complete theory of quantum gravity would have) at a UV scale $M$ lead to large corrections that spoil slow-roll inflation. In \cite{Kaloper:2011jz}, the authors were especially interested in the viability of inflation with a quadratic potential, so the emphasis was on ensuring that the corrections to the quadratic potential were small.  Corrections of the form
\be
	\Delta \CL = (F^{(4)})^2 \sum_k \frac{(F^{(4)})^{2k}}{M^{4k}}\ ,
\ee
lead, after integrating out $F$, to corrections to the classical scalar potential $V_c = \half \mu^2\phi^2$ of the form
\be
	\delta V = V_c \sum_k \frac{V_c^k}{M^{4k}}
\ee
This was based on the assumption that the additional UV degrees of freedom at scale $M$ couple most naturally to $F$ rather than to ${\tilde F}$.  If they coupled instead to $\tF$, then we would find
\be
	\delta V = V \sum_k \left(\frac{q}{M^2}\right)^2 \frac{V^k}{M^{4k}}\ .
\ee
Since the model is viable only if $q \ll M^2$ to begin with, such corrections are further suppressed.  The upshot is that we find corrections to $V \sim \half m^2\phi^2$ when ${\tilde F} \gg (M^2 g^2, M^4)$, depending on whether the UV degrees of freedom at scale $M$ couple to $F$ or to ${\tilde F}$, respectively. On the other hand, instability to domain wall nucleation becomes dangerous when ${\tilde F} \gg M_T^4$, the scale set by the domain wall tension.  A regime in which $V(\phi)$ deviates from a quadratic potential without rapid membrane nucleation occurring requires $M_T \gg M$ if UV degrees of freedom couple to ${\tilde F}$, and $M_T^4 \gg q M^2$ if the UV degrees of freedom couple to $F$.  The latter case gives a slightly wider range for $M_T$. An obvious way for this to happen is for $F$ to couple to relatively light moduli; it was argued in \cite{Dong:2010in}\ that this would generically lead to flattening.

An analogy to the massive Schwinger model arises if we promote the 2d $\theta$ parameter to a dynamical scalar field $\chi$ with periodicity $2\pi$, and consider the 2d Maxwell field as the analog of ${\tilde F}$.  The fermions, dual to the Sine-Gordon scalar field, are the analogs of the domain walls in four dimensions.   If the canonically normalized field ${\tilde \chi}$ has dimensionless radius $R$, then we find that the physical mass will satisfy $m_{\chi}^2 = e^2/R^2$, in analogy to the condition  (\ref{eq:percond}).  In the discussions in \S2.2-2.3, the massive fermions also provide the additional UV degrees of freedom at mass $M = m$. In this case we find, roughly, that the "tension" of the 0d domain walls is $M_T = m$ as well, which is hardly a surprise since the "domain walls" and the fundamental degrees of freedom are the same when $e^2 \ll m^2$.  Thus, as discussed in \S2.3.1, a perturbatively stable branch with an energy that is nonquadratic in $\theta$, requires coupling $F_{\mu\nu}$ to light neutral fields $\psi$, via couplings such as $f(\psi) F^2$.

As discussed in the introduction, a related 4d field theory model is a an axion coupled to the topological charge $\tr F \wedge F$ on a nonabelian gauge theory. A specific, nonsupersymmetric, strongly-coupled version was studied in \cite{Dubovsky:2011tu}.  In that example the monodromy, the flattening of the potential, and the instability were all generated by the underlying nonabelian gauge dynamics.  The flattening, in particular, is associated with a reduction of the mass gap as a function of $\theta$ along a given branch.  One might ask whether a 2d analog with a dynamically generated mass would lead to a similar effect.  We now turn to a canonical example in this class.

%

\section{The ${\mathbb CP}^N$ model}

In this section we wish to study a theory for which the theta term arises in an asymptotically free theory with a dynamically generated mass scale.  An obvious candidate is the ${\mathbb CP}^N$ model, long studied as a 2d analog to QCD.  The large N limit provides a potential analog to the existing work in 4 dimensions on $\theta$-angle monodromy \cite{Witten:1979vv,Witten:1980sp,Witten:1998uka,Dubovsky:2010je,Dubovsky:2011tu}, and places the theory under computational control.  At low energies, the model is described by an Abelian gauge field coupled to massive charged bosons, with the gauge coupling and boson mass generated dynamically.  With all of that, we will find that the regime in which $\CE(\theta)$ deviates from $\CE \sim \theta^2$ is identical to the regime when pair production of charged bosons becomes unsuppressed.  Nonetheless, the detailed calculations are interesting in this case.  For example, the dynamically generated mass depends on the $\theta$-induced electric field; furthermore, we find a barest hint of $\CE(\theta)$ beginning to flatten before the theory becomes unstable.  

\subsection{Introduction to the model}

The two-dimensional nonlinear $\sigma$-model with target space ${\mathbb CP}^N$ model can be written as \cite{DAdda:1978uc,Witten:1978bc}:
\be
	S = \int d^2 x \left[\frac{N}{g^2} |(\d - i A) z^i|^2 - \lambda\left(z^i z_i^* - 1\right) + \frac{\theta}{2\pi} \eps^{\mu\nu}\d_{\mu} A_{\nu}\right]
\ee
Here $z^{i -1,\ldots, N}$ are a set of $N$ complex scalar fields; $\lambda$ is a Lagrange multiplier enforcing $\sum_i |z^i|^2 = 1$, and $A_{\mu}$ is a nondynamical gauge field which gauges away an overall phase rotation of $z^i$.  This combination of restricting to $S^{2N - 1} \subset \mathbb{C}^N$ followed by the gauging is equivalent to the description of $\cpn$ as ${\mathbb{C}}^N/{\mathbb{C}}^*$.  Upon integrating out $A$, the $\theta$ term is equivalent to $\theta \int {}^*\omega$ where $\omega$ is the Kahler form of $\cpn$.

The effective action about $E = \p_0 A_1 - \p_1 A_0  = 0$ was computed in \cite{DAdda:1978uc,Witten:1978bc}.  We will compute the effective action $E$ increases. We do this integrating out $z^i$, $\lambda$. The effective action upon integrating out $z^i$ is
\be
	S_{eff} = i N \tr \ln \left( - (\d - i A)^2 - \frac{\lambda g^2}{N}\right) + i \int d^2 x \left[ \lambda + \frac{\theta}{2\pi} \eps^{\mu\nu}\d_{\mu} A_{\nu}\right]
\ee
Following \cite{Schwinger:1951nm}, the effective Lagrangian is:
\be\label{eq:cpneffact}
	L_{eff} = - \frac{i N }{4\pi} \int_0^{\infty} \frac{ds}{s} \frac{E}{\sinh Es} e^{-i \frac{\lambda g^2 s}{N}} + \lambda + \frac{\theta E}{2\pi}
\ee
This is divergent at $s\to 0$; $s$ has dimensions of $(length)^2$, so the divergence is quadratic.

\subsection{The dynamically generated mass gap}

The Lagrange multiplier $\lambda$ couples as a mass term to the bosons $z^i$.  At leading order in $1/N$, $\lambda$ acquires a nonvanishing expectation value, found by solving for $d_{\lambda} L = 0$:
\be\label{eq:finitegap}
	1 - \frac{g^2}{4\pi} \int_0^{\infty} \frac{E ds}{\sinh s} e^{-i M^2 s} = 0
\ee
where I have used $M^2 = g^2\lambda/N$.  Let us first discuss the $E \to 0$ limit \cite{ Witten:1978bc}:
\be
	1 - \frac{g^2}{4\pi} \int_0^{\infty} \frac{ds}{s} e^{-i M_0^2 s} = 0
\ee
This is logarithmically divergent.  In order to more easily do the integral, we can analytically continue $s = -i t$, and cut off the integral over the {\it Euclidean} proper time at $t = 1/\Lambda^2$, where $\Lambda$ is the UV cutoff.  This gives:
\be
	1 - \frac{g^2}{4\pi} \int_{1/\Lambda^2}^{\infty} \frac{dt}{t} e^{- M_0^2 t} = 0
\ee
This integral is dimensionless.  The leading divergence is $-\ln M^2/\Lambda^2$.  There is no finite piece; the remaining terms are powers of $M^2/\Lambda^2$, which we will ignore.  Thus, we find
\be
	\frac{\lambda g^2}{N} \equiv M_0^2 = \Lambda^2 \exp\left\{- \frac{4\pi}{g^2} - \gamma\right\}
\ee
as in \cite{Witten:1978bc}; here $\gamma$ is the Euler-Mascheroni constant.  If we consider the low-energy effective action for small fluctuations of $z, A$, this becomes the dynamically induced mass for $z^i$.

Now let us consider $E \neq 0$, in the regime $E/\Lambda^2 \ll 1$, $E/M^2 \ll 1$. The finite part of the integral in (\ref{eq:finitegap}) can be computed exactly: 
\begin{eqnarray}
	I & = & \int_0^{\infty} ds \left(\frac{E}{\sinh E s} - \frac{1}{s} \right) \cos M^2 s \nonumber\\
	& = & - \Re \psi \left( \half + \frac{i M^2}{2 E} \right) + \ln \left(\frac{M^2}{2 E}\right)
		- 2\pi i \frac{e^{-\pi M^2/E}}{1 + e^{-\pi M^2/E}}\ ,
\end{eqnarray}
where $\psi(x)$ is the digamma function. In the limit $M^2 \ll E$, the imaginary part is exponentially small and we will ignore it in solving the gap equation.  We can also compute the finite part of (\ref{eq:finitegap}) in a power series by expanding $1/\sinh$ in a power series.  This also misses the imaginary part, which arise from the poles in the integrand of (\ref{eq:finitegap}) along the imaginary axis. The leading terms in the finite part of the gap equation are:
\be
	1 + \frac{g^2}{4\pi} \left( \gamma + \ln \left(\frac{M^2}{\Lambda^2} \right) - \frac{E^2}{6 M^4} - \frac{7 E^4}{60 M^8} + \ldots\right) = 0
\ee
We can solve for $M^2$ in a power series:
\be\label{eq:perturbedgap}
	M^2 = M_0^2 + \frac{E^2}{6 M_0^2} + \frac{3 E^4}{40 M_0^6} + \ldots
\ee
Thus $M^2$ depends explicitly on $E$.  

\subsection{Effective action}

Let us return to (\ref{eq:cpneffact}).  This contains both a quadratic and logarithmic divergence in the $E \to 0$ limit:
\be
	L_{div} = - \frac{i N}{4\pi} \int_{\eps}^{\infty} \frac{ds}{s^2} e^{- i M^2 s}
\ee
we set $\eps = \frac{1}{i\Lambda^2}$.  If we redefine $i s = t$, then the lower limit becomes the standard proper time cutoff.  If we throw out all terms which are positive powers of $\eps$ (these will be of order $M^2/\Lambda^2$), then:
\be
	L_{div} = \frac{N \Lambda^2}{4\pi} + \frac{N M^2}{4\pi} \left(\gamma - 1 + \ln \left(\frac{M^2}{\Lambda^2}\right)\right)
\ee
I will ignore the first quadratic divergence, which renormalizes the cosmological constant.

The finite part of $L_{eff}$ can be calculated perturbatively in $E^2/M^4$ by: redefining the integration variable in (\ref{eq:cpneffact}) as $u = M^2 s$, expanding the hyperbolic cosecant in a power series, and subtracting the leading term which gave the UV divergence.  As before, this procedure will miss effects nonperturbative in $|E|/M^2$: in particular the imaginary part will not appear.  The finite part computed in this way is a power series in $E^2/M^4$ identical to the finite part of Eq. (24) in \cite{Blau:1988iz} (after Lorentzian continuation).  If we include the first two terms in this expansion, we find that:
\be
	L_{eff} = L_{div} + \frac{N M^2}{g^2} + \frac{N E^2}{24\pi M^2} + \frac{7 N E^4}{720 \pi M^6}
\ee
Finally, if we insert (\ref{eq:perturbedgap}), the leading terms in an expansion of the effective action in $E^2/M_0^4$ is:
\be\label{eq:effactseries}
	L_{eff} = \frac{N E^2}{24 \pi M_0^2} + \frac{N E^4}{160 \pi M_0^6} - \frac{\theta E}{2\pi}
\ee
The first term was also found in \cite{Witten:1978bc}; it is a dynamically generated kinetic term for the gauge field.  We are also left with charged bosons, which have an effective Lagrangian
\be
	L_{bos} = \frac{N}{g^2} \sum_i |(\d - i A) z^i |^2 - \frac{N M^2(E)}{g^2} \sum_i |z^i|^2
\ee
Note that this leads to an effective coupling for the gauge field which depends on $|z|$.

\subsection{Theta dependence}

Next, we wish to find the potential energy for the theory as a function of $\theta$; we do this by computing the Hamiltonian.  If we fix the gauge $A_0 = 0$, the canonical momentum for $A_1$ is:
\be
	\Pi = \frac{N E}{12\pi M_0^2} + \frac{N E^3}{40 \pi M_0^6} - \frac{\theta}{2\pi}
\ee
We solve for $E/M_0^2$ in a power series in $x = \frac{12\pi}{N}\left(\Pi - \frac{\theta}{2\pi}\right)$, to find
\be
	\frac{E}{M_0^2} = x - \frac{3}{10} x^3
\ee
Finally, computing the Hamiltonian density, we find:
\be
	H = \frac{N M_0^2}{24\pi} \left[\frac{12\pi}{N} \left(\Pi - \frac{\theta}{2\pi}\right)\right]^2
		 - \frac{N M_0^2}{160 \pi} \left[\frac{12\pi}{N} \left(\Pi - \frac{\theta}{2\pi}\right)\right]^4 + \ldots
\ee
Since the $U(1)$ is compact in this model, $\Pi$ is quantized. If we set $\Pi = 0$, it will be fixed (until pairs of charged bosons with mass $M$ nucleate.) The potential energy becomes a function of $x = 6 \theta/N$:
\be\label{eq:thetapot}
	V(\theta) = \frac{N M_0^2}{24\pi} \left[ \left(\frac{6\theta}{N}\right)^2 - \frac{3}{20} \left(\frac{6\theta}{N}\right)^4 + \ldots \right] = N M_0^2 \CE(\frac{6 \theta}{N}) \ .
\ee
It is interesting to note that the correction flattens the potential slightly. Note that this is somewhat analogous to the functional form $V(\theta) = N^2 {\cal V}\left(\frac{\lambda\theta}{N}\right)$ found in four-dimensional theories \cite{Witten:1979vv,Witten:1980sp,Witten:1998uka,Dubovsky:2011tu}.  However, as the only dimensionless coupling is absorbed into the dynamical mass $M_0$ via dimensional transmutation, and there is no other scale in the problem (so long as we stay at energies well below the cutoff $\Lambda$) there is no additional dimensionless coupling that appears in (\ref{eq:thetapot}).  

\subsection{Stability and flattening}

In this theory we have a set of charged bosons with mass $M^2$, which can screen the electric field via pair creation.  The probability of pair production should be
\be
	P \propto e^{ - \pi M(E)^2/\Delta V}
\ee
where $M$ is the mass of the boson, and $\Delta V$ is the difference in potential energy between the exterior and interior of the boson-antiboson pair. It is easy to show that for large $N$, $\theta/N \sim 1$, $\Delta V = E$. Let the Lagrangian have the form 
\be
	L(E) = N \ell(E) - \frac{\theta}{2\pi} E\ .
\ee
Then the canonical momentum is
\be
	P = N \ell'(E) - \frac{\theta}{2\pi}
\ee
If we consider a fixed branch $P = 0$ of the monodromy potential, then 
\be\label{eq:efortheta}
	N \ell'(E) = \frac{\theta}{2\pi}\ . 
\ee
Now if $\theta \to \theta - 2\pi$, (\ref{eq:efortheta}) implies that $E \to E - \delta E$ where
$N \ell''(E) \delta E = - 1$; thus we can work to first order in $\delta E$ in the large $N$ limit. The variation of the Hamiltonian is
\begin{eqnarray}
	\delta H & = & \delta \left(\frac{\theta}{2\pi} E - N \ell(E)\right)\nonumber\\
	& = & E + \delta E\left(\frac{\theta}{2\pi} - N \ell'(E)\right) = E
\end{eqnarray}
The decay rate for boson-antiboson pairs is then
\be
	\Gamma \sim N e^{- \pi M^2(E)/|E|}
\ee
where the prefactor arises from the number of bosons that could be produced.   This probability becomes appreciable when $|E|/M^2$ is of order $1/\ln N$, which occurs before $E(\theta)$ begins to deviate appreciably from being quadratic.  Again, the essential point is that as we adiabatically increase $\theta$, the instanton action is $S \sim N/|\theta|$ and the effective potential is a power series in $\theta/N$.  There is no additional parameter that might allow for a separation between these regimes, unlike the case of \cite{Dubovsky:2011tu}.

\section{Conclusions}

Another model one could explore is the sigma model on the Grassmannian $U(n+m)/U(n)\times U(m)$.  This can be written as a $U(m)$ gauge theory coupled to $n$ charged bosons. There is a $\theta$ angle for the $U(1)$ factor; and the Maxwell term and boson mass are generated dynamically\cite{Macfarlane:1979hi,Brezin:1980ms}. (See also vol. II of \cite{Deligne:1999qp}).  I leave this for future work.

More generally, it would be nice to have a deeper understanding of the fact that all of the models here become unstable just as, if not before, $\CE(\theta)$ deviates from quadratic, distinct from the example in \cite{Dubovsky:2011tu}  As I stated in the introduction, a part of the explanation could be that the 2d $\theta$ term always couples to an abelian factor.

\vskip .5cm

\centerline{\bf Acknowledgements}
\vskip .3 cm
I would like to thank Sergei Dubovsky, Matthew Headrick, Jonathan Heckman, Shamit Kachru, Nemanja Kaloper, Matthew Kleban, Matthew Roberts, and Howard Schnitzer for useful discussions and helpful comments.  I would also like to thank the Center for Cosmology and Particle Physics at NYU, the Aspen Center for Physics, and the Stanford Institute for Theoretical Physics for their hospitality at crucial times during this project.  My research is supported by DOE Grant DE-FG02-92ER40706.

\eject
\bibliographystyle{utphys}
\bibliography{dlrrefs}

\end{document}